\documentstyle[12pt]{article}

\begin{document}

\newcommand{\reell}
{{\kern+.25em\sf{R}\kern-.78em\sf{I}\kern+.78em\kern-.25em}}
\newcommand{\komplex}
{{\sf{C}\kern-.46em\sf{I}\kern+.46em\kern-.25em}}
\newcommand{\posganz}
{{\kern+.25em\sf{N}\kern-.86em\sf{I}\kern+.86em\kern-.25em}}
\newcommand{\ganz}
{{\kern+.25em\sf{Z}\kern-.78em\sf{Z}\kern+.78em\kern-.65em}}
\newcommand{\Hamilton}
{{\kern+.25em\sf{H}\kern-.86em\sf{I}\kern+.86em\kern-.25em}}
\newcommand{\Cayley}
{{\sf{O}\kern-.56em\sf{I}\kern+.56em\kern-.25em}}
\newcommand{\unit}
{{\sf{1}\kern-.18em\sf{I}\kern+.18em\kern-.18em}}
\newcommand{\opunit}
{{\sf{1}\kern-.29em\sf{1}\kern+.29em\kern-.33em}}
\newcounter{blabla}
\newcommand{\be}{\begin{equation}}
\newcommand{\ee}{\end{equation}}
\newcommand{\bdm}{\begin{displaymath}}
\newcommand{\edm}{\end{displaymath}}
\newcommand{\beann}{\begin{eqnarray*}}
\newcommand{\eeann}{\end{eqnarray*}}
\newcommand{\bea}{\begin{eqnarray}}
\newcommand{\eea}{\end{eqnarray}}

\newcommand{\nn}{\nonumber \\}

\newcommand{\koszul}{\bigtriangledown}
\newcommand{\artref}[4]{{\sc #1} {\it #2} {\bf #3} #4}
\newcommand{\tartref}[5]{{\sc{#1}},{#2},{\it{#3}}{\ \bf{#4}},{#5}}
\newcommand{\bookref}[2]{{\sc #1}, #2}
\newtheorem{definition}{Definition}[section]
\newtheorem{exercise}{Exercise}[section]
\newtheorem{question}{Question}[section]
\newtheorem{lemma}[definition]{Lemma}
\newtheorem{proposition}[definition]{Proposition}
\newtheorem{theorem}[definition]{Theorem}
\newtheorem{corollary}[definition]{Corollary}

\newcommand{\resection}[1]{\setcounter{equation}{0}\section{#1}}
\newcommand{\appsection}{\addtocounter{section}{1}
	   \setcounter{equation}{0}\section*{Appendix\Alph{section}}}
\renewcommand{\theequation}{\arabic{equation}}

\begin{titlepage}

\def\mytoday#1{{ } \ifcase\month \or
 January\or February\or March\or April\or May\or June\or
 July\or August\or September\or October\or November\or December\fi
 \space \number\year}
\noindent
\hspace*{11cm} BUTP--94/5\\
\vspace*{1cm}
\begin{center}
{\LARGE Gauge-Symmetry Breakdown at the Horizon of Extreme Black Holes}

\vspace{1cm}

J. Bi\v{c}\'{a}k
\\
Department of Theoretical Physics \\
Charles University \\
V Hole\v{s}ovi\v{c}k\'{a}ch 2 \\
18000 Prague 8, Czech Republic	\\
and \\
Institute for Theoretical Physics \\
University of Bern \\
Sidlerstrasse 5, CH-3012 Bern, Switzerland
\\ \vspace{0.5cm}

C. Cris, P. H\'{a}j\'{\i}\v{c}ek and A. Higuchi
\\
Institute for Theoretical Physics \\
University of Bern \\
Sidlerstrasse 5, CH-3012 Bern, Switzerland
\\ \vspace{0.5cm}

\mytoday \\ \vspace*{0.5cm}

\nopagebreak[4]

\begin{abstract}
Static solutions of the Einstein-Yang-Mills-Higgs system containing extreme
black holes are studied. The field equations imply strong restrictions on
boundary values of all fields at the horizon. If the Yang-Mills radial electric
field $E$ is non-zero there, then all fields at the horizon take values in the
centralizer of $E$. For the particular case of SU(3), there are two different
kinds of centralizers: two-dimensional abelian (Cartan subalgebra) and
four-dimensional (su(2)$\times$u(1)) ones. The two-dimensional centralizer
admits only constant fields: even the geometry of the horizon is that of
constant curvature. If the cosmological constant $\Lambda$ is negative, a
two-surface of any genus is possible; for positive curvature, only spherically
symmetrical horizons are allowed. For the four-dimensional centralizer, all
spherically symmetrical horizons are explicitly given.

Finally, some complete spacetime solutions are constructed whose horizons have
the
structure found by our method. There are also abelian solutions of a new type.
In some cases there are different spacetimes having the same type of horizon.

\end{abstract}

\end{center}

\end{titlepage}

	 \section{Introduction and Summary}

	 Classical solutions to Einstein-Yang-Mills equations have
	 attracted much attention recently. Spherically symmetrical
	 nonabelian solutions (so-called ``solitons'') have been found
	 (\cite{1}) and studied (\cite{2,3}). It is remarkable that
	 gauge fields and gravitational singularities can cancel and
	 finite-energy, regular solutions exist. It is hoped that such
	 non-perturbative effects may have some consequences on
	 quantum level as well. Similar (but singular) solutions that
	 contain event or Killing horizons (``black holes'') also
	 exist (\cite{4,5,6}). For reviews, see
	 \cite{7} and most recently \cite{8}.

	 The existence of such solutions reveals an unexpected
	 richness of the Einstein-Yang-Mills system. In particular,
	 the non-abelian black holes represent counterexamples to the
	 ``no hair conjecture''. In addition, there are speculations
	 about the role that these solutions could play in
	 microphysics (\cite{9,10}).

	 There has been much interest in the extreme
	 Reissner-Nordstr\"{o}m black holes within the standard
	 Einstein-Maxwell theory. They admit surprisingly simple
	 solutions of the perturbation equations \cite{11};
	 some of them seem to be stable with respect to both classical
	 and quantum processes and there are attempts to interpret
	 them as solitons (\cite{7,12}); also, they admit
	 supersymmetry (\cite{13}). Very recently extreme
         Reissner-Nordstr\"{o}m black holes were discussed in the context
	 of the Einstein-Maxwell theory with a cosmological constant
	 $\Lambda$ \cite{14}. In particular, it is possible to analyze
	 collisions of black holes analytically and study the cosmic
	 censorship hypothesis by considering charged black
	 holes with $\Lambda \neq 0$ (\cite{15,16}).

	 The geometry of extreme black hole spacetimes exhibits some
	 characteristic features. There is no Einstein-Rosen bridge (or
	 ``wormhole'') joining two asymptotically flat regions and
	 containing a minimal 2-sphere. Instead, there is one
	 asymptotically flat and one asymptotically cylindrical region
	 on each $t =$ const. hypersurface. We call the boundary of
	 the cylindrical region ``internal infinity''. Such an
	 internal infinity is a compact two-dimensional spacelike
	 surface. The hypersurfaces $t =$ const. do not intersect the
	 horizon, but only approach such an intersection at the
	 internal infinity.

	 These properties yield a tool for a systematic study of
	 extreme black holes (\cite{17,18}). As the space
	 approaches a cylinder at an internal infinity, all radial
	 derivatives of smooth fields must vanish there, and the field
	 equations degenerate in the limit to a system of some
	 algebraical, some first-order and some second-order
	 differential equations on a compact two-dimensional surface;
	 the second order equations are elliptic. Such a system has
	 few solutions. For example, there are only spherically
	 symmetrical solutions in most cases \cite{19}.

	 In the present paper, we start an investigation of the
	 structure of internal infinity for the
	 Einstein-Yang-Mills-Higgs system with the gauge group SU(3).
	 We choose the Higgs field with values in the adjoint
	 representation of the group and with the usual simple
	 biquadratic potential. Other potentials (see, e.g. \cite{20})
	 could be analyzed similarly and will lead to analogous
	 results.

	 The plan of the paper is as follows.

	 In Section 2, we write down the Lagrangian, explain the
	 character of the limit to the internal infinity and write
	 down the limiting system of equations. In Section 3, we
	 observe that a non-zero ``electric'' charge $q$ of the black
	 hole leads to a kind of ``symmetry breaking'': the values of
	 all fields must lie in the centralizer of $q$.  All gauge
	 non-equivalent charges lie in a sector of the two-dimensional
	 Cartan subalgebra. The direction at the boundary of this sector has a
	 four-dimensional non-abelian centralizer corresponding to the
	 subgroup U(1)$\times$SU(2) of SU(3). For all other directions
	 in the sector, the centralizers coincide with the
	 two-dimensional Cartan subalgebra itself, so the solution is
	 necessarily abelian.

	 In Section 4, we find all solutions for the two-dimensional
	 centralizers. There are only abelian Higgs vacuum (true or
	 false) fields with maximal symmetry. For negative values of
	 the cosmological constant there are internal infinities of
	 all orientable topologies (arbitrary genus) as the curvature
	 scalar can be negative. If the curvature is positive, a
	 spherically symmetrical solution results.

	 We also find some complete
	 spacetime solutions with these boundaries. The corresponding
         gepmetry is the Reissner-Nordst\"{o}m spacetime. The gauge field
	 is characterized by {\em four} gauge invariant charges (two
	 electric and two magnetic). We observe that the known
         Reissner-Nordstr\"{o}m ``embeddings'' (\cite{21,22}) have
	 at most two charges (electric and magnetic).

	 In Section 5, we study the spherically symmetrical solutions
	 on the four-dimen\-sion\-al centralizer. As SU(3) is broken to
	 U(1)$\times$SU(2), we can use the well-known SU(2) ansatz. We
	 find two classes of solutions. 1) The Higgs field has its
	 true vacuum value and its component into the ``hypercharge'' $(T_8)$
	 direction does not vanish. This leads to the
         Reissner-Nordstr\"{o}m spacetime with a magnetic SU(2)-charge as
	 well as an electric and magnetic ``hypercharge''. 2) Only the
	 SU(2)-part of the Higgs field is non-zero, but smaller
	 than the true vacuum value. We show that these are the
	 internal infinities either to the countable family of extreme
	 black holes described in \cite{23} or to those found in
	 \cite{24}.

	 In the Appendix it is indicated how the existence of extreme
	 black hole solutions leads to restrictions of the ranges of
	 the free parameters in the Lagrangian of the model.

	 In this way all static ``electric'' solutions (non-zero
	 electric charge) are likely to be found: it is very plausible
	 that static solutions have maximal symmetry and higher genus
	 may lead to only  asymptotically locally anti-de Sitter
         spacetimes. Moreover, Reissner-Nordstr\"{o}m-type solutions are
	 expected to be stable, in contrast to more complicated solutions.
	 Further study is necessary especially for
	 ``magnetic'' solutions (zero electric charge).

	 Our conventions: the metric signature is $+2$, the Riemann
	 tensor $R^\mu_{~\nu \rho \sigma}$ is defined by
	 $$
	 R^\mu_{~\nu \rho \sigma} = \Gamma^\mu_{\nu \sigma, \rho} -
	 \Gamma^\mu_{\nu \rho, \sigma} + \ldots ,
	 $$
	 and the Ricci tensor $R_{\mu \nu}$ is
	 $$
	 R_{\mu \nu} = R^\rho_{~\mu \rho \nu}.
	 $$
	 The units are chosen such that $\hbar = c = 1$.

	 \section{The Model}

	 \subsection{The Lagrangian}

	 We consider a four-dimensional manifold ${\cal M}$ and a
	 system of fields consisting of a spacetime metric $g_{\mu
	 \nu}$, a Yang-Mills field $W_\mu$ and a Higgs scalar field
	 $Q$; $W_\mu$ and $Q$ have their values in the Lie algebra $l{\cal G}$
	 of a simple, matrix, compact Lie group ${\cal G}$.

	 The Lagrangian is given by
	 \be
	 L = L_G + L_M,
	 \ee
	 where
	 \bea
	 L_G& =& - \frac{1}{16 \pi \gamma^2} \int d^3x \sqrt{-^4g}\; (^4R-
	 2 \Lambda), \nonumber \\
	 L_M &=& \frac{1}{4 \pi} \int d^3x \sqrt{-^4g} \left
	 [\frac{1}{4e^2} (G_{\mu \nu}, G^{\mu \nu}) + \frac{1}{2}
	 (D_\mu Q, D^\mu Q) + V(Q) \right], \nonumber
	 \eea
	 $\gamma^2$ is the Newton constant ($\gamma$ is the Planck
	 length),
	 $$
	 ^4g = \det (g_{\mu \nu}),
	 $$
	 $^4R$ is the curvature scalar of $g_{\mu\nu}$, $\Lambda$ is the
	 cosmological constant, $e$ is the coupling constant of the
	 Yang-Mills field,
	 $$
	 G_{\mu \nu} = \partial_\mu W_\nu - \partial_\nu W_\mu - [W_\mu,
	 W_\nu]
	 $$
	 is the gauge field strength, (.,.) is an invariant positive
	 definite quadratic form on $l{\cal G}$
	 $$
	 (\cdot , \cdot) = - \frac{1}{v} (\cdot , \cdot)_K,
	 $$
	 proportional to the Killing form $(\cdot ,\cdot)_K$ in such a way
	 that most of the structure constants of an orthonormal basis are small
	 integers,
	 $$
	 D_\mu Q = \partial_\mu Q - [W_\mu, Q]
	 $$
	 is the covariant derivative defined by the gauge field,
	 $$
	 V(Q) = \frac{1}{8} k [(Q, Q) - F^2]^2
	 $$
	 is the Higgs potential and $k$ and $F$ are positive
	 constants. Under a gauge transformation $U(x), x \in {\cal
	 M}, U \in {\cal G}$, the fields transform as follows:
	 \bea
	 W'_\mu &=& U^{-1} W_\mu U - U^{-1} \partial_\mu U,
	 \nonumber \\
	 Q'& =& U^{-1} Q U.  \nonumber
	 \eea
	 The field equations resulting from the Lagrangian (1) read
	 \be
	 R_{\mu\nu} - \frac{1}{2}\; {^4R} g_{\mu \nu} = 8 \pi \gamma^2
	 \bar{T}_{\mu \nu},
	 \ee
	 \be
	 (-^4g)^{-1/2} D_\nu \left[(-^4g)^{1/2} G^{\mu \nu} \right] +
	 e^2 [D^\mu Q, Q] = 0,
	 \ee
	 \be
	 (-^4g)^{-1/2} D_\nu \left[ (-^4g)^{1/2} D^\nu Q\right] -
	 \frac{k}{2} \left[ (Q,Q) - F^2 \right] Q = 0,
	 \ee
	 where
	 \be
	 \bar{T}_{\mu \nu} = T_{\mu \nu} -  \frac{\Lambda}{8 \pi
	 \gamma^2} g_{\mu \nu}
	 \ee
	 and
	 \bea
	 T^{\mu \nu} &=& \frac{1}{4 \pi e^2} \left[ (G^{\mu \rho},
	 G^\nu_\rho) - \frac{1}{4} g^{\mu \nu} (G^{\rho \sigma},
	 G_{\rho \sigma})\right] +  \nonumber \\
	 &&+ \frac{1}{4 \pi} \left [ (D^\mu Q, D^\nu Q) - \frac{1}{2}
	 g^{\mu \nu} (D_\rho Q, D^\rho Q) - g^{\mu \nu} V (Q) \right
	 ].
	 \eea

	 \subsection{The Ansatz for the Solution}

	 We will look for solutions $({\cal M}, g, W, Q)$ to eqs.
	 (2)--(6) describing a static extremal black hole. Thus, we
	 will assume the following properties.

	 1. $({\cal M}, g, W, Q)$ is static: there is a timelike
	 vector field $\xi^\mu (x)$ with the norm $N(x)$,
	 $$
	 N^2 = - g_{\mu \nu} \xi^\mu \xi^\nu,
	 $$
	 and a map $X : {\cal M} \rightarrow l{\cal G}$ such that
	 \bea
	 {\cal L}_\xi g_{\mu \nu} &=& 0 \nonumber \\
	 {\cal L}_\xi W_\mu &=& [W_\mu, X] - \partial_\mu X,\nonumber \\
	 {\cal L}_\xi Q& =& [Q,X], \nonumber
	 \eea
	 where ${\cal L}_\xi$ is the Lie derivative with respect to
	 $\xi^\mu (x)$.

	 2. Let $\Sigma$ be an inextendable $\xi$-orthogonal
	 hypersurface in $({\cal M}, g)$. Then $\Sigma$ is a Cauchy
	 hypersurface for $({\cal M}, g)$. Further, $\Sigma$ is
	 complete with respect to the distance function associated
	 with the positive definite metric induced on $\Sigma$ by
	 $g_{\mu \nu}$.

	 3. The function $N(x)$ on $\Sigma$ is positive,
	 $$
	 N(x) > 0 \hspace{0.3cm} \forall x \in \Sigma
	 $$
	 not bounded below away from zero. There is $\epsilon > 0$ such
	 that each component ${\cal K}$ of the set
	 $$
	 N(x) < \epsilon
	 $$
	 in $\Sigma$ satisfies the conditions

	 (i) $N(x)$ has no critical points in ${\cal K}$,

	 (ii) the surfaces $N(x) = \mbox{const.}$ are compact.

	 Thus ${\cal K}$ is diffeomorph to $\reell \times {\cal H}$, where
	 ${\cal H}$ is a compact two-dimensional manifold and there are
	 coordinates $N, x^2, x^3$ in ${\cal K}$ such that the metric has the
	 form
	 \be
	 ds^2 = - N^2 dt^2 + \rho^2 dN^2 + g_{AB} dx^A dx^B,
	 \ee
	 $A,B = 2,3$, and $\rho$ and $g_{AB}$ depend only on $N$ and
	 $x^A$.

	 Each component ${\cal K}$ contains an internal infinity as $N(x)
	 \rightarrow 0$. As $\Sigma$ is complete, the distance of any
	 point of $\Sigma$ to the infinity must diverge. Thus
	 $$
	 \int^{a}_{0} \rho d N = \infty,
	 $$
	 if $a < \epsilon$, and we must have
	 $$
	 \lim_{N \rightarrow 0} \rho^{-1} = 0.
	 $$

	 4. Let us choose a particular gauge in each ${\cal K}$:

	 \begin{description}

	\item (i) The gauge frame is parallel along the $N$-curves in
	 $\Sigma$:
	 $$
	 W_1 = 0.
	 $$

	\item (ii) The gauge frame is propagated by the static group along
	 the orbits of $\xi^\mu$:
	 $$
	 \partial_0 W_\mu = \partial_0 Q = 0.
	 $$
	 Then, the functions $\rho$, $g_{AB}$, $W_0$, $W_A$ and $Q$ of
	 the variables $N, x^2, x^3$ are continuous in ${\cal K}$ together
	 with all their derivatives up to order 2. The function
	 $$
	 (N\rho)^{-1} \partial_1 W_0
	 $$
	 is continuous together with its all first order derivatives.
	 The limits $N \rightarrow 0$ of all these functions and
	 derivatives exist and are smooth functions of $x^2$ and
	 $x^3$.

	 \end{description}

	 This condition of regularity of the fields at the
	 internal infinity is an essential part of the ansatz. It is a
	 technical point: physically, one can only justify the
	 regularity at all points of the spacetime, in particular of
	 the horizon. Some examples show that the regularity at the
	 horizon does not imply the regularity at the internal
	 infinity (\cite{13}).

	 The limits $N \rightarrow 0$ of the fields $g_{AB}$, $W_0$,
	 $W_A$ and $Q$ form fields on the manifold ${\cal H}$. Eqs.
	 (2)-(6) imply a system of equations for these fields on
	 ${\cal H}$ that has been derived \cite{17}. In the present
	 special case these equations take the following form (we
	 omit the limit sign):
	 \be
	 \frac{1}{(N\rho)^2} = \gamma^2 \left [ \frac{(E,E)}{e^2} +
	 \frac{(B,B)}{e^2} - 2 V - \frac{\Lambda}{\gamma^2} \right
	 ],
	 \ee
	 \be
	 R = 2 \gamma^2 \left[ \frac{(E,E)}{e^2} + \frac{(B,B)}{e^2} +
	 g^{AB} (D_A Q, D_B Q) + 2 V + \frac{\Lambda}{\gamma^2}
	 \right],
	 \ee
	 \be
	 \frac{(E,E)}{e^2} + \frac{(B,B)}{e^2} - 2 V = \mbox{const.},
	 \ee
	 \bea
	 (D_AQ, D_B Q) &=& \frac{1}{2} g_{AB} g^{KL} (D_K Q, D_L Q),\\
	 D_A E &=& 0, \\
	 \left[ E,Q \right] &=& 0,
	 \eea
	 \be
	 \frac{\epsilon^{AB}}{\sqrt{g}} D_B B + g^{AB} e^2 [Q, D_B Q]
	 = 0,
	 \ee
	 \be
	 \frac{1}{\sqrt{g}} D_A (\sqrt{g} g^{AB} D_B Q) -
	 \frac{\partial V}{\partial Q} = 0
	 \ee
	 where $E$ and $B$, given by
	 $$
	 G_{10} = N \rho E, \hspace{0.2cm}  G_{AB} = \epsilon_{AB}
	 \sqrt{g} B
	 $$
	 are radial electric and magnetic fields on ${\cal H}$, $R$ is the
	 curvature scalar of $g_{AB}$,
	 $$
	 g = \det (g_{AB}),
	 $$
	 $\epsilon_{AB}$ and $\epsilon^{AB}$ are the antisymmetric
	 densities, and the internal index at $\partial V/\partial Q$
	 is raised by the metric ($\cdot , \cdot$).

	 The strucutre of eqs. (8)--(15) leads naturally to two
	 different sorts of solutions: those with $E \neq 0$ at least
	 at one point of ${\cal H}$, and those with $E=0$ everywhere
	 on ${\cal H}$. The former solutions are called
	 ``electric'', the latter ``magnetic''. The ``symmetry
	 breaking'' is a property of the electric solutions.

	 \section{Electric Solutions}

	 \subsection{The Role of Centralizers}

	 Suppose that there is a point	$p \in {\cal H}$ such that
	 $E(p) \neq 0$. Then eq. (12) implies that there is a gauge
	 along ${\cal H}$ (consisting possibly of more patches) such
	 that $E$ is a fixed constant element of $l{\cal G}$ everywhere along
	 ${\cal H}$ (equal to $E(p)$). Let us choose this gauge. Then,
	 $$
	 \partial_A E = 0,
	 $$
	 and eq. (12) implies
	 \be
	 [W_A, E ] = 0.
	 \ee
	 It follows, in particular, that
	 \be
	 [B,E] = 0.
	 \ee
	 Let us denote by $Z_X$ the centralizer of the element $X$ in
	 the Lie algebra $l{\cal G}$:
	 $$
	 Z_X = \left \{ Y \in l{\cal G} \mid [Y,X] = 0 \right \}.
	 $$
	 Then, eqs. (16), (17) and (13) imply the following
	 proposition.

	 \vspace{0.2cm}

	 \noindent

	 {\bf Proposition:} For the above choice of gauge along ${\cal
	 H}$,
	 the values of all fields $W_A, E, B$ and $Q$ lie in $Z_E$.

	 Thus, centralizers will play a crucial role for the
	 electric solutions: the gauge group will be ``broken'' to a
	 subgroup which is generated by a centralizer.

	 \subsection{General Properties of Centralizers}

	 Let us collect those properties of centralizers that hold
	 generally for simple groups ${\cal G}$ and that will be relevant to
	 our problem. (They might be already given in the literature;
	 we were not able to find a suitable reference. However, the
	 proofs are short and a mere list of the properties, which
	 would be necessary in any case, is not much shorter than our
	 Sections 3.2 and 3.3.)

	 Directly from the definition, we have
	 $$
	 aX \in Z_X \hspace{0.3cm} \forall a \in \reell,
	 $$
	 (so $Z_X$ is always at least one-dimensional), and
	 $$
	 Z_{aX} = Z_X \hspace{0.3cm} \forall a \in \reell.
	 $$
	 Thus, it is sufficient to look for normalized elements of
	 $l{\cal G}$:
	 $$
	 (X,X) =1.
	 $$
	 Then the elements $X$ and $-X$ of this Killing sphere
	 $S$ have the same centralizer.

	 Suppose that
	 $$
	 Y = ad(g) X, \hspace{0.3cm} g \in {\cal G}.
	 $$
	 Then,
	 $$
	 Z_Y\; \;  = \; \; ad(g) Z_X.
	 $$
	 This follows immediately from the relation:
	 $$
	 [ad(g) U, ad(g) V] = ad(g) [U,V].
	 $$

	 The adjoint representation of ${\cal G}$ on $l{\cal G}$ leaves the
	 Killing sphere $S$ invariant. Centralizers of two elements
	 of the
	 same orbit of ${\cal G}$ in $S$ can be mapped on each other by
	 $ad(g)$ for some $g \in {\cal G}$; the corresponding solutions of
	 eqs. (8)-(15) will, therefore, be gauge related. However,
	 elements from different orbits will give gauge non-equivalent
	 solutions. Thus, it is sufficient for our purposes to study
	 only some typical representatives of the orbits. Let us show
	 how this can be done for SU(3).

	 \subsection{The Case of SU(3)}

	 Let us choose the following basis for the Lie algebra su(3)
	 (Gell-Mann matrices times ``$i$''):

	 $$
	 T_1 = \left ( \begin{array}{rrr}
	 0&i&0 \\ i&0&0 \\ 0&0&0 \end{array} \right ), \hspace{0.3cm}
	 T_2 = \left( \begin{array}{rrr}
	 0 & 1 & 0 \\ -1 & 0 &0 \\ 0 & 0 & 0 \end{array}
	 \right),
	 $$

	 $$
	 T_3 = \left ( \begin{array}{rrr}
	 i&0&0 \\ 0&-i&0 \\ 0&0&0 \end{array} \right ), \hspace{0.3cm}
	 T_4 = \left( \begin{array}{rrr}
	 0 & 0 & i \\ 0 & 0 &0 \\ i & 0 & 0 \end{array} \right),
	 $$

	 $$
	 T_5 = \left ( \begin{array}{rrr}
	 0&0&1 \\ 0&0&0 \\ -1&0&0 \end{array} \right ), \hspace{0.3cm}
	 T_6 = \left( \begin{array}{rrr}
	 0 & 0 & 0 \\ 0 & 0 &i \\ 0 & i & 0 \end{array}
	 \right),
	 $$

	 $$
	 T_7 = \left ( \begin{array}{rrr}
	 0&0&0 \\ 0&0&1 \\ 0&-1&0 \end{array} \right ), \hspace{0.3cm}
	 T_8 = \frac{1}{\sqrt{3}} \left( \begin{array}{rrr}
	 i & 0 & 0 \\ 0 & i &0 \\ 0 & 0 & -2i \end{array}
	 \right),
	 $$

	 In this basis, the form ($\cdot , \cdot$) is given by
	 $$
	 (X,Y) = X_a Y_a.
	 $$
	 Observe that the diagonal basis elements $T_3$ and $T_8$ span
	 the Cartan subalgebra $C$ of su(3).

	 Let $X \in l{\cal G}$. As $X$ is skew-symmetric, there is an
	 element $U \in$ SU(3) such that
	 $$
	 U^{-1} X U = ad (U) X
	 $$
	 is a diagonal matrix, or
	 $$
	 ad(U) X \in C.
	 $$
	 Hence: each orbit of ${\cal G}$ in $l{\cal G}$ has a
	 representative in $C$, and we can restrict ourselves to
	 centralizers of the elements of $C \cap S$.

	 Suppose that $X, Y \in C$, and that there is $U\in$ SU(3) such
	 that
	 $$
	 Y = U^{-1} X U.
	 $$
	 Then $Y$ must have the same eigenvalues as $X$. As both
	 matrices are diagonal, they can differ only in the order of
	 their diagonal elements. Moreover, we can achieve any
	 permutation of any diagonal elements of any $X
	 \in C$ by the corresponding permutation matrix $U$ (an
	 orthogonal matrix with 0 and 1 as elements). For even
	 permutations, $U\in$ SU(3); for odd permutations, $- U \in$
	 SU(3). Hence exactly those elements of $C$ that differ by
	 a permutation of their diagonal elements lie in the same orbit
	 of SU(3) in su(3).

	 Let us introduce the coordinates $\rho$ and $\alpha$ on $C$
	 by
	 $$
	 X  = T_3 \rho \sin \alpha + T_8 \rho \cos \alpha.
	 $$
	 For $X \in C \cap S$, $ \rho = 1$. The six permutations of
	 the diagonal elements of $X$ leave $\rho$ invariant and send
	 $\alpha$ to
	 $$
	 \begin{array}{rrrrrrrrr}
	& \alpha, &\alpha &+& \frac{2 \pi}{3}&, \hspace{0.2cm}
	 &\alpha &+& \frac{4
	 \pi}{3}, \\
	 &- \alpha, &- \alpha &+& \frac{2 \pi}{3}&,
	 \hspace{0.2cm} &- \alpha &+& \frac{4
	 \pi}{3}.\end{array}
	 $$
	 Hence for each orbit of SU(3) in su(3), there is exactly one
	 representative in the sector
	 \be
	 0 \leq \alpha < \frac{\pi}{3}.
	 \ee
	 Finally, we must find the centralizers to all elements in the
	 sector (18). This is easy: all matrices $Y$ which commute
	 with a given diagonal matrix $X$ have a block-diagonal form:
	 the non-diagonal elements $Y_{ij}$ must vanish, if
	 $$
	 X_{ii} \neq X_{jj}.
	 $$
	 Thus, all $X$'s with three different diagonal elements
	 commute only with diagonal matrices. This is the case for
	 $$
	 0 < \alpha < \frac{\pi}{3},
	 $$
	 and $Z_X = C$ in this case. Only for $\alpha = 0$ --
	 $T_8$ direction, $X_{11} = X_{22}$, so $Z_X =$
	 su(2)$\times$u(1). Hence, we have a two-dimensional abelian
	 centralizer $C$ if $0< \alpha < \frac{\pi}{3}$, generating
	 the subgroup U(1)$\times$U(1), and
	 a four-dimensional non-abelian centralizer if $\alpha = 0$,
	 generating the subgroup SU(2)$\times$U(1).

	 \section{Two-Dimensional Centralizers}

	 \subsection{Solutions at the Internal Infinity}

	 Let us assume that
	 $$
	 E = E_\alpha T_\alpha \neq 0,
	 $$
	 where greek indices run through two values: 3 and 8. All
	 different solutions correspond to the range
	 $$
	 0 < \frac{E_3}{E_8} < \sqrt{3}.
	 $$
	 Then, we must have
	 $$
	 \begin{array}{lll}
	 W_A & = & W^\alpha_A T_\alpha, \\
	 Q & = & Q_\alpha T_\alpha, \\
	 B & = & B_\alpha T_\alpha,
	 \end{array}
	 $$
	 and eqs. (8)-(15) become

	\setcounter{equation}{18}

	 \be
	 \frac{1}{(N\rho)^2}  =  \gamma^2 \left[ \frac{E_\alpha
	 E_\alpha}{e^2} + \frac{B_\alpha B_\alpha}{e^2} - \frac{k}{4}
	 (Q_\alpha Q_\alpha - F^2)^2 - \frac{\Lambda}{\gamma^2} \right
	 ],
	 \ee
	 \be
	 R  =  2 \gamma^2 \left[ \frac{E_\alpha
	 E_\alpha}{e^2} + \frac{B_\alpha B_\alpha}{e^2} + \frac{k}{4}
	 (Q_\alpha Q_\alpha - F^2)^2
	 + g^{AB} \partial_A Q_\alpha
	 \partial_B Q_\alpha + \frac{\Lambda}{\gamma^2} \right],
	 \ee
	 \be
	  \frac{E_\alpha
	 E_\alpha}{e^2} + \frac{B_\alpha B_\alpha}{e^2} - \frac{k}{4}
	 (Q_\alpha Q_\alpha - F^2)^2 = \mbox{const.},
	 \ee
	 \bea
	 \partial_A Q_\alpha \partial_B Q_\alpha &=& \frac{1}{2} g_{AB}
	 g^{KL} \partial_K Q_\alpha \partial_L Q_\alpha
	 \\
	 \partial_A E_\alpha &=& 0,
	 \\
	 \partial_A B_\alpha &=& 0,
	 \eea
	 \be
	 \frac{1}{\sqrt{g}} \partial_A (\sqrt{g} g^{AB} \partial_B
	 Q_\alpha) - \frac{k}{2} (Q_\beta Q_\beta - F^2) Q_\alpha =
	 0.
	 \ee
	 We obtain immediately
	 \be
	 E_\alpha = \mbox{const.} \; , \hspace{0.33cm} B_\alpha =
	 \mbox{const.}
	 \ee
	 Then eq. (21) yields
	 $$
	 Q^2_3 + Q^2_8 = \mbox{const.}
	 $$
	 This means that the function $Q_3(x^2,x^3)$ is not
	 independent of $Q_8(x^2,x^3)$, and so the matrix on the
	 l.h.s. side of eq. (22) is degenerate. However, it must be
	 proportional to $g_{AB}$ which is non-degenerate. Hence the
	 proportionality factor must be zero, and we obtain
	 \be
	 Q_\alpha = \mbox{const.}
	 \ee
	 Substituting eq. (27) in eq. (25), we find that
	 $$
	 (Q_\beta Q_\beta - F^2) Q_\alpha = 0.
	 $$
	 Hence there are only two cases:
	 $$
	 \begin{array}{ll}
	 \mbox{A)}& \hspace{0.3cm} Q_\alpha = 0, \\
	 \mbox{B)}& \hspace{0.3cm} Q^2_3 + Q^2_8 = F^2,
	 \end{array}
	 $$
	 A) being the ``false'' and B) the ``true'' Higgs vacuum. The
	 remaining eqs. (19) and (20) become
	 \bea
	 \frac{1}{(N\rho)^2} & = & \gamma^2 \left( \frac{E_\alpha
	 E_\alpha}{e^2} + \frac{B_\alpha B_\alpha}{e^2} - \epsilon
	 \frac{kF^4}{4} - \frac{\Lambda}{\gamma^2} \right ), \\
	 R & = & 2 \gamma^2 \left( \frac{E_\alpha
	 E_\alpha}{e^2} + \frac{B_\alpha B_\alpha}{e^2} + \epsilon
	 \frac{kF^4}{4} + \frac{\Lambda}{\gamma^2} \right ),
	 \eea
	 where $\epsilon = 1$ for the case A and $\epsilon = 0$ for
	 the case B. Eq. (29) implies that $({\cal H}, g_{AB})$ is a
	 compact two-dimensional space of constant curvature. If $R
	 >0$, $({\cal H}, g_{AB})$ is a sphere of radius $r_H$,
	 $$
	 R = \frac{2}{r^2_H}.
	 $$
	 If $R= 0$, we will have toroidal topology, and if $R <0$,
	 the genus can have any value larger than 1. However, $R \leq
	 0$ only if $\Lambda < 0$. The corresponding black hole
	 spaces are then likely to be only  asymptotically locally
	 anti-de Sitter (with some identifications made at anti-de Sitter
	 infinity). This question must be further studied.

	 Let us limit ourselves to spherical holes. If we denote by
	 $q_\alpha$ the electric and by $p_\alpha$ the magnetic
	 charges of the hole, then
	 $$
	 E_\alpha = \frac{q_\alpha}{r^2_H} \hspace{0.3cm},
	 \hspace{0.3cm} B_\alpha = \frac{p_\alpha}{r^2_H}.
	 $$
	 Let us choose $x^2 = \theta$ and $x^3 = \phi$, the ordinary
	 spherical coordinates on ${\cal H}$.
	 Then, we can set
	 $$
	 W^\alpha_\theta = 0, \hspace{02cm} W^\alpha_\phi = - p_\alpha \cos
	 \theta.
	 $$
	 Observe that eq. (29) implies then an analogy to the
	 ``extremality'' condition
	 $$
	 m^2 = q^2 + p^2,
	 $$
	 namely
	 $$
	 r^2_H = \gamma^2 \left ( \frac{q_\alpha q_\alpha}{e^2} +
	 \frac{p_\alpha p_\alpha}{e^2} \right ) + \left ( \Lambda +
	 \frac{\epsilon \gamma^2 k F^4}{4} \right) r^4_H.
	 $$
	 In the case A, different solutions can be parametrized by
	 the angle $\alpha$ between the vector $q_\alpha$ and hypercharge
	 axis, and by three further parameters: lengths of the vectors
	 $q_\alpha$ and $p_\alpha$, and the angle $\beta$ between $q_\alpha$
	 and $p_\alpha$. In the case B, the solutions are classified by the
	 same angle $\alpha$, and by four further parameters: three as the
	 case A, and the angle between $q_\alpha$ and $Q_\alpha$.

	 This type of solution has been described for any gauge group
	 in [17].

	 \subsection{Spacetime Solutions}

	 Let us look for a whole asymptotically flat spacetime with
	 the fields $W_\mu$ and $Q$, whose internal infinity has the
	 structure found in the previous section. To this end, we
	 choose the following ansatz:

	 \begin{itemize}

	 \item[1)] spherical symmetry everywhere,

	 \item[2)] the fields $W_\mu$ and $Q$ take values from $C$
	 everywhere,

	 \item[3)] the Higgs field $Q$ is in its false (case A) or true
	 (case B) vacuum state every\-where.

	 \end{itemize}

	 The assumptions 2 and 3 for $Q$ lead to a) vanishing of the
	 $Q$-part in $T^{\mu \nu}$ with the exception of the term
	 $$
	 - \frac{1}{4 \pi} V (Q) g^{\mu \nu} = - \frac{k \epsilon}{32
	 \pi} F^4 g^{\mu \nu}
	 $$
	 (cf. eq. (6)), b) eq. (4) being identically satisfied, and c)
	 vanishing of the source term in eq. (3).

	 The Yang-Mills field splits into two ``Maxwell fields''
	 $W_\mu^3$ and $W_\mu^8$, and eq. (3) becomes
	 \be
	 \partial_\nu \left [ (-^4g)^{1/2} G^{\mu \nu}_3 \right] = 0,
	 \ee
	 \be
	 \partial_\nu \left [ (-^4g)^{1/2} G^{\mu \nu}_8 \right] = 0,
	 \ee
	 where
	 $$
	 G^\alpha_{\mu \nu} = \partial_\mu W^\alpha_\nu - \partial_\nu
	 W^\alpha_\mu \; \; , \; \; \alpha = 3,8.
	 $$

	 Finally, Eq. (2) becomes
	 \be
	 R_{\mu \nu} - \frac{1}{2}\; {^4R} g_{\mu \nu} + \bar{\Lambda}
	 g_{\mu \nu} = 8 \pi \gamma^2 (T^3_{\mu \nu} + T^8 _{\mu
	 \nu}),
	 \ee
	 where the ``effective'' cosmological constant $\bar{\Lambda}$
	 is given by
	 $$
	 \bar{\Lambda} = \Lambda + \frac{1}{4} \epsilon k \gamma^2
	 F^4. $$

	 The solution to the system (30)-(32) is just the
         Reissner-Nordstr\"{o}m spacetime with the metric
	 \be
	 ds^2 = - \phi dt^2 + \phi^{-1} dr^2 + r^2 d \Omega^2,
	 \ee
	 where
	 \be
	 \phi(r) = 1 - \frac{2 \xi}{r} + \frac{\eta^2}{r^2} -
	 \frac{1}{3} \zeta r^2,
	 \ee
	 $\xi, \eta$ and $\zeta$ being some constants. An extreme
	 horizon will exist only if the function $\phi r^2$ has a
	 double-root $r_H$. This happens if and only if the following
	 two relations are satisfied
	 \be
	 \zeta r^4_H - r^2_H + \eta^2 = 0,
	 \ee
	 \be
	 \xi = r_H (1 - \frac{2}{3} \zeta r^2_H).
	 \ee
	 Then
	 \be
	 r^2 \phi (r) = \left ( - \frac{1}{3} \zeta r^2 - \frac{2}{3}
	 \zeta \; r_H \;r- \zeta r^2_H + 1 \right ) \left (r- r_H \right)^2,
	\ee
	 and we have two cases

	 (i) Cosmological double horizon:
	 $$
	 \zeta \in \left( 0, \frac{1}{4\eta^2} \right),
	 $$
	 \be
	 r_H = \frac{1}{\sqrt{\zeta}} \sqrt{\frac{1}{2} +
	 \frac{1}{2} \sqrt{1-4 \zeta \eta^2}},
	 \ee
	 \be
	 \xi = r_H \left ( \frac{2}{3} - \frac{1}{3} \sqrt{1-4 \zeta
	 \eta^2} \right ).
	 \ee
	 The spacetime is not static near this double horizon so that the
	 method of this paper does not seem applicable.

	 (ii) Black-hole double horizon:
	 $$
	 \zeta \in \left(- \infty, \frac{1}{4\eta^2}\right),
	 $$
	 \be
	 r_H = \frac{\mid \eta \mid}{\sqrt{\frac{1}{2} + \frac{1}{2}
	 \sqrt{1 - 4 \zeta \eta^2}}},
	 \ee
	 \be
	 \xi = r_H \left ( \frac{2}{3} + \frac{1}{3} \sqrt{1-4 \zeta
	 \eta^2} \right ).
	 \ee
	 For $\zeta \leq 0$, only black-hole double horizon can arise.
	 Consider $\zeta >0$. Then it is easy to see that the function
	 $r^2 \phi (r)$, if $\eta \neq 0$, must always have one
	 negative root and one largest postive root $r_c$. In general,
	 it may also have two different real roots, $r_1$, $r_2$ such
	 that $0 < r_1 < r_2 < r_c$; $r_1, r_2$ and $r_c$ correspond
	 then to a single inner, outer and cosmological horizon
	 respectively. If
	 $$
	 0 < r_1 = r_2 = r_H < r_c,
	 $$
	 we have the black-hole double horizon.
	 If $0 < r_1 < r_2 = r_c = r_H$, we use the term ``cosmological''
	 double horizon.
	 Clearly, both black-hole
	 double horizon and cosmological double horizon can never
	 occur simultaneously. For the black-hole double horizon
	 $$
	 (r^2 \phi)''\mid_{r = r_H} \geq 0
	 $$
	 so that $r_H < \frac{1}{\sqrt{2 \zeta}}$. For the cosmological
	 double horizon,
	 $$
	 (r^2 \phi)''\mid_{r_H} < 0
	 $$
	 and
	 $$
	 r_H > \frac{1}{\sqrt{2\zeta}}.
	 $$
	 A special case arises when
	 $$
	 r_1 = r_2 = r_c = \frac{1}{\sqrt{2 \zeta}}.
	 $$
	 Then all three horizons coincide,
	 $$
	 (r^2 \phi)'' \mid_{r_H} = 0,
	 $$
	 $$ \eta^2 = \frac{9}{8} \xi^2 \; , \; \zeta = \frac{1}{4
	 \eta^2}.
	 $$
	 In \cite{25}, some conformal diagrams corresponding to the
	 double
         horizons are drawn. Very recently, special Reissner-Nordstr\"{o}m
	 spacetimes with the cosmological double horizons were
	 analyzed in connection with a possible violation of the
	 cosmic censorhsip in \cite{16}. In our case, the
	 non-vanishing cosmological constant -- which plays a crucial
	 role in \cite{16} -- can entirely arise from the Higgs field.

	 Eq. (29) will be satisfied, if
	 $$
	 \zeta = \Lambda + \frac{1}{4} \epsilon \gamma^2 k F^4 =
	 \bar{\Lambda}
	 $$
	 and
	 $$
	 \eta^2 = \frac{\gamma^2}{e^2} (q_\alpha q_\alpha + p_\alpha
	 p_\alpha).
	 $$
	 Comparing eqs. (7) and (33), we obtain
	 $$
	 N = a \sqrt{\phi},
	 $$
	 $$
	 \rho^2 = \frac{1}{\phi N'^{2}} = \frac{4}{a^2 \phi'^{2}},
	 $$
	 and
	 $$
	 \lim_{N \rightarrow 0}(N\rho)^2 = \lim_{r \rightarrow r_H}
	 \frac{4 \phi}{\phi'^{2}},
	 $$
	 where $a$ is an arbitrary positive constant. Setting
	 $$
	 \phi = H(r) (r-r_H)^2,
	 $$
	 we have
	 $$
	 \lim_{N \rightarrow 0} (N\rho)^2 = \frac{1}{H(r_H)},
	 $$
	 and eq. (39) yields
	 $$
	 H(r_H) = \frac{1-2 \zeta r^2_H}{r^2_H}.
	 $$
	 With this value of $(N\rho)^{-2}$, eq. (28) is identically
	 satisfied. Hence our solution at the internal infinity
	 corresponds to the given spacetime solution, as claimed.

	 \section{Four-Dimensional Centralizer}

	 \subsection{Solutions at the Internal Infinity}

	 The four-dimensional centralizer is spanned by the basis
	 elements $T_1, T_2, T_3$ and $T_8$. We assume, therefore, the
	 fields of the form
	 \bea
	 W^a_0 & = & (0, \ldots, 0, W^8_0), \\
	 W^a_1 & = & 0, \nonumber \\
	 W^a_A & = & (W^1_A, W^2_A, W^3_A, 0, \ldots, 0, W^8_A),
	 \nonumber \\
	 E_a & = & (0, \ldots,0 , E_8), \\
	 B_a & = & (B_1, B_2, B_3, 0, \ldots, 0, B_8), \nonumber \\
	 Q_a & = & (Q_1, Q_2, Q_3, 0, \ldots, 0, Q_8).\nonumber
	 \eea
	 Thus the fields split in a U(1) field $W^8_\mu$, $E_8$,
	 $B_8$, and $Q_8$, and an SU(2) field $W^k_\mu, E_k = 0, B_k$,
	 and $Q_k$. We will use the latin indices $a, b, c$ for the
	 whole field, i.e., they assume the values  $1, \ldots, 8$;
	 the indices $i,j,k$ will take only the values $1,2,3$ and
	 will distinguish the components of the SU(2) field. We will also
	 use the vector notation $\vec{W}_\mu, \vec{E}, \vec{B}$ and
	 $\vec{Q}$ for the SU(2) field.

	 Eqs. (12) to (15) split also into U(1) equations -- for the
	 8th components, and SU(2) equations -- for the first 3 components.

	 Non-trival U(1) equations:
	 \bea
	 \partial_A E_8 & = & 0, \\
	 \partial_A B_8 & = & 0, \\
	 \frac{1}{\sqrt{g}} \partial_A \left ( \sqrt{g} g^{AB}
	 \partial_B Q_8 \right) & = & \frac{1}{2} k \left ( \vec{Q}^2
	 + Q^2_8 - F^2 \right) Q_8.
	 \eea

	 Nontrivial SU(2) equations:
	 \be
	 \frac{\epsilon^{AB}}{\sqrt{g}} \bigtriangledown_B \vec{B} +
	 2 e^2 g^{AB} \vec{Q} \times \bigtriangledown_B \vec{Q} = 0,
	 \ee
	 \be
	 \frac{1}{\sqrt{g}} \bigtriangledown_A \left ( \sqrt{g} g^{AB}
	 \bigtriangledown_B \vec{Q} \right) - \frac{1}{2} k \left (
	 \vec{Q}^2 + Q^2_8 -F^2\right) \vec{Q} = 0,
	 \ee
	 \be
	 \vec{B} = \frac{1}{\sqrt{g}} \epsilon^{AB} \left (\partial_A
	 \vec{W}_B + \vec{W}_A \times \vec{W}_B \right),
	 \ee
	 where we have introduced the SU(2) covariant derivative,
	 $$
	 \bigtriangledown_A \vec{X} = \partial_A \vec{X} + 2 \vec{W}_A
	 \times \vec{X}.
	 $$
	 The components $a = 4,5,6,7$ of all eqs. (12)-(15) are
	  satisfied identically.

	 The remaining eqs. (8)-(11) take the form
	 \bea
	 \frac{1}{(N\rho)^2} & = & \gamma^2 \left [ \frac{1}{e^2} (E^2_8 +
	 B^2_8) + \frac{1}{e^2} \vec{B}^2 - \right. \nonumber \\
	 && \left.- \frac{1}{4}k ( \vec{Q}^2 + Q^2_8 - F^2)^2 -
	 \frac{\Lambda}{\gamma^2} \right],
	 \eea
	 \bea
	 R &=& 2 \gamma^2 \left [ \frac{1}{e^2} (E^2_8 + B^2_8) +
	 \frac{1}{e^2} \vec{B}^2 + \frac{\Lambda}{\gamma^2}
	 \right. \nonumber \\
	 &&\left.+ g^{AB} (\bigtriangledown_A \vec{Q} \cdot
	 \bigtriangledown_B
	 \vec{Q} + \partial_A Q_8 \partial_B Q_8) + \frac{1}{4} k
	 (\vec{Q}^2 + Q^2_8 - F^2 )^2 \right],
	 \eea
	 \be
	 \frac{1}{e^2} (E^2_8 + B^2_8) + \frac{1}{e^2} \vec{B}^2 -
	 \frac{1}{2} k (\vec{Q}^2 + Q^2_8 - F^2)^2 = \mbox{const.},
	 \ee
	 and
	 \bea
	 &&\bigtriangledown_A \vec{Q} \cdot \bigtriangledown_B \vec{Q} +
	 \partial_A Q_8 \partial_B Q_8 = \nonumber \\
	 &&=  \frac{1}{2} g_{AB} g^{KL} (\bigtriangledown_K \vec{Q} \cdot
	 \bigtriangledown_L \vec{Q} + \partial_K Q_8 \partial_L Q_8).
	 \eea

	 In contrast to the case of the two-dimensional (abelian)
	 centralizers treated in the previous section, it does not now
	 appear easy to deduce from the system of eqs. (44)--(53) that
	 the internal infinity must necessarily be a compact space of
	 constant curvature. Therefore, we make a spherically
	 symmetric {\em ansatz} for both the metric and the fields at
	 ${\cal H}$. We thus assume
	 \be
	 g_{AB} dx^A dx^B = r^2_H (d \theta^2 + \sin^2 \theta d
	 \varphi^2),
	 \ee
	 so that
	 $$
	 R= \frac{2}{r^2_H}.
	 $$
	 The spherically symmetric ansatz for the SU(2) fields (see,
	 e.g. \cite{20,26}) is given by
	 \bea
	 \vec{W}_r &=& 0, \nonumber \\
	 \vec{W}_\theta & = & w (\sin \varphi, - \cos \varphi, 0),
	 \nonumber \\
	 \vec{W}_\varphi & = & w \sin \theta (\cos \theta \cos
	 \varphi, \cos \theta \sin \varphi, - \sin \theta) ,  \\
	 \vec{Q} & = & {\cal Q} (\sin \theta \cos \varphi, \sin \theta
	 \sin \varphi, \cos \theta),
	 \eea
	 where $w$ and ${\cal Q}$ are constants. Substituting for
	 $\vec{W}_A$ into eq. (49), we obtain
	 \be
	 \vec{B} = b (\sin \theta \cos \varphi, \sin \theta \sin
	 \varphi, \cos \theta),
	 \ee
	 where
	 \be
	 b = 2 w (w -1) r^{-2}_H.
	 \ee
	 The spherically symmetric ansatz for the U(1)-fields is
	 simply
	 $$
	 E_8 = \mbox{const.}\;,\; B_8 = \mbox{const.},
	 $$
	 $$
	 Q_8 = \mbox{const.}
	 $$

	 Let us turn to eqs. (44)--(53) and substitute our ansatz into
	 them. Eqs. (44) and (45) are satisfied identically. Eq. (46)
	 admits only the following two types of solutions:
	 \be
	 Q_8 \neq 0  \; \;, \; \; \vec{Q}^2 + Q^2_8 = F^2,
	 \ee
	 and
	 \be
	 Q_8 = 0.
	 \ee
	 We shall discuss the two solutions separately.

	 \subsubsection*{A) $Q_8 \neq 0,\; \; \vec{Q}^2 + Q^2_8 = F^2$}

	 This is the ``true Higgs vacuum'', $V(Q)=0$. Using eqs. (55)
	 and (56), we find eq. (48) to be satisfied only if
	 \be
	 w = \frac{1}{2}.
	 \ee
	 This, however, implies that
	 \be
	 \bigtriangledown_A \vec{Q} = \bigtriangledown_A \vec{B} = 0,
	 \ee
	 so that eqs. (47) and (53) are also satisfied. Clearly the
	 same is true with eq. (52) since $\vec{B}^2 = b^2 = $ const.
	 and $\vec{Q}^2 = {\cal Q}^2 =$ const. Denoting the electric
	 and magnetic hypercharges by $q$ and $p$, we have
	 \be
	 E_8 = \frac{q}{r^2_H} \; \; \; , \; \; \; B_8 =
	 \frac{p}{r^2_H},
	 \ee
	 and in view of eqs. (58) and (61), we can rewrite eqs. (50)
	 and (51) in the form
	 \bea
	 \left( \frac{\gamma}{N \rho}\right)^2 & =
	 &\frac{\gamma^4}{e^2
	 r^4_H} (q^2 + p^2 + \frac{1}{4}) -\gamma^2 \Lambda, \\
	 \left ( \frac{\gamma}{r_H} \right)^2 &= &\frac{\gamma^4}{e^2
	 r^4_H} (q^2 + p^2 + \frac{1}{4}) + \gamma^2 \Lambda.
	 \eea

	 The magnitudes of the Higgs fields $\vec{Q}^2 = {\cal Q}^2$
	 and $Q_8$ are only constrained by
	 $$
	 {\cal Q}^2 + Q^2_8 = F^2.
	 $$
	 Hence, these solutions can be described by three parameters:
	 $q,p$ and ${\cal Q}$.

	 \subsubsection*{B) $Q_8 = 0$}

	 Starting again from the ansatz (55) and (56) (and assuming in
	 general non-vanishing $E_8$ and $B_8$), we find in this case
	 that eq. (47) yields the relation
	 \be
	 b = - 2 e^2 {\cal Q}^2,
	 \ee
	 if $w \neq \frac{1}{2}$. As a consequence of eq. (48) we
	 obtain
	 \be
	 \frac{1}{4} \frac{k r^2_H}{(1-2 w)^2} (F^2 - {\cal Q}^2) = 1.
	 \ee
	 Therefore, solutions exist only if
	 \be
	 F^2 \geq {\cal Q}^2.
	 \ee
	 Expressing $w$ from eq. (67), one finds
	 \be
	 w = \frac{1}{2} \pm \frac{1}{4} \sqrt{k} r_H \sqrt{F^2- {\cal
	Q}^2}.
	 \ee
	 Clearly, the equality in (68) can arise only for $w =
	 \frac{1}{2}$ (cf. (67)), which is the case of the true Higgs
	 vacuum discussed above. Hereafter we assume that $F^2 > {\cal
	 Q}^2$.

	 It can be easily checked that eq. (53) is satisfied as well
	 as the condition (52). The remaining eqs. (50) and (51) can
	 be rewritten into the form
	 \bea
	 \left (\frac{\gamma}{N\rho} \right)^2 & = &
	 \frac{\gamma^4}{e^2} (E^2_8 + B^2_8) +
	 4 e^2 (\gamma {\cal Q})^4 - \frac{1}{4} k \left[(\gamma F)^2
	 - (\gamma {\cal Q})^2 \right]^2 - \gamma^2 \Lambda, \\
	 \left (\frac{\gamma}{r_H} \right)^2 & = &
	 \frac{\gamma^4}{e^2} (E^2_8 + B^2_8) +
	  \frac{1}{4} (16e^2 - k) (\gamma {\cal Q})^4 + \frac{1}{4} k
	 (\gamma F)^4 + \gamma^2 \Lambda.
	 \eea
	 These represent restrictions on the parameters. All possible
	 classes of the parameters are briefly mentioned in the
	 Appendix.

	 Summarizing, the solutions (55), (56), (57) with $b$ and $w$
	 given by eqs. (66) and (69), with in general non-vanishing
	 $E_8$ and $B_8$, represent a gravitating
      't~Hooft-Polyakov-type monopole in the SU(2) fields which
	 has additional electric and magnetic U(1) fields in the 8th direction
	 and an extreme black hole ``in the middle''. However, these fields
	 are defined at the internal infinity only, and it remains to
	 be seen whether they can be extended to global spacetime
	 solutions.

	 \subsection{Spacetime Solutions}

	 As in the case of the two-dimensional centralizers, we
	 are going to construct some solutions in the whole spacetime
	 that, at the internal infinity, go over to the fields found
	 in Section 5.1.

         \subsubsection*{I. Solutions of the Reissner-Nordstr\"{o}m Type}
	 \subsubsection*{Ia. The True-Higgs-Vacuum Solutions}

	 Assume everywhere in a static spherically symmetric spacetime
	 that

	 (i) the Yang-Mills field $\vec{W}_k$ has the form (55) with
	 $w = \frac{1}{2}$, and $\vec{W}_0 = 0$;

	 (ii) the Higgs field $\vec{Q}$ is given by eq. (56) with
	 ${\cal Q} =$ const.;

	 (iii) the field $Q_8 =$ const. satisfying together with
	 $\vec{Q}$ the true vacuum condition:
	 $$
	 {\cal Q}^2 + Q^2_8 = F^2,
	 $$

	 (iv) the U(1) field $W^8_0$ has the form
	 \be
	 W^8_0 = - \frac{c}{r},
	 \ee
	 where $c=$ const. is to be determined by the condition
	 $$
	 E_8 = \frac{1}{N \rho} \partial_r W^8_0
	 $$
	 and
	 \be
	 W^8_2 = 0 \; , \; W^8_3 = - p \cos \theta.
	 \ee
	 Under these conditions it is easy to see that $D_\mu Q = 0$
	 everywhere. Since, in addition, the Higgs field is in the
	 true vacuum state, it neither contributes to $T^{\mu
	 \nu}$ (eq. (6)), nor to the source term in the Yang-Mills eq.
	 (3), and eq. (4) for the Higgs field is identically satisfied.
	 The gauge fields split everywhere into SU(2) and U(1) fields, and
	 eq. (3) becomes
	 \bea
	 \frac{1}{\sqrt{- ^4g }} \partial_\nu \left ( \sqrt{- ^4g}
	  G^{\mu \nu}_i \right) & - & [W_\nu , G^{\mu \nu}_i] = 0
	 \\
	 \partial_\nu \left( \sqrt{- ^4g} G^{\mu \nu}_8 \right) & = &
	 0.
	 \eea
	 Under our assumptions for $W^a_\mu$, one easily finds the
	 components of $G^a_{\mu \nu}$, the only non-vanishing ones being
	 \bea
	 \vec{G}_{\theta \varphi} &= & - \frac{1}{2} \sin \theta (\sin
	 \theta \cos \varphi, \sin \theta \sin \varphi, \cos \theta),
	 \\
	 G^8_{tr} & = & \frac{c}{r^2} \; \; , \; \; G^8_{\theta
	 \varphi} = p \sin \theta.
	 \eea
	 The form of the sources indicates that the metric will again
         be of the Reissner-Nordstr\"{o}m type. Indeed, assuming the
	 metric (33), we have $\sqrt{- ^4g} = r^2 \sin \theta$,
	 and one can immediately check that the Yang-Mills eqs. (74), (75) are
	 satisfied by the fields (76) and (77) everywhere.

	 Hence, we can proceed as in Section 4.2 and write the
	 function $\phi(r)$ in the form (34) and find again the
	 expressions for the location $r_H$ of the black hole or
	 cosmological double horizons and the mass parameters $\xi$.
	 Putting then
	 \bea
	 \eta^2 & = & \left( \frac{\gamma}{e} \right)^2 (q^2 + p^2 +
	 \frac{1}{4}), \\
	 \zeta & = & \Lambda,
	 \eea
	 we find eqs. (65) and (35) to be satisfied. Finally, we can
	 check that eq. (64) is also satisfied in an analogous way as
	 in Section 4.2. By using eqs. (72), (73) and the expression
	 for $N\rho$ found at the end of Section 4.2,
	 we obtain
	 $$
	 E_8 = \frac{\sqrt{ 1- 2 \Lambda r^2_H}}{r^3_H} c
	 $$
	 so that the constant $c$ appearing in eqs. (72) and (77) is
	 given by
	 \be
	 c = \frac{r_H}{\sqrt{1-2 \Lambda r^2_H}}q.
	 \ee
	 To summarize, the global solution is of the
         Reissner-Nordstr\"{o}m-de Sitter form (33) and (34) with $\eta$
	 and $\zeta$ given by eqs. (78) and (79), $\xi$ and $r_H$ by
	 eqs. (38), (39), (40) and (41) in Section 4.2. The Yang-Mills
	 fields are determined by eqs. (55), (61), (72), (73), (76), (77)
	 and (80), the Higgs field by eqs. (56) with ${\cal Q} =$
	 const. and by $Q_8 =$ const. such that ${\cal Q}^2 + Q^2_8 =
	 F^2$.

	 Setting $q = p = Q_8 = 0$, we obtain the field of a pure
	 magnetic monopole with the magnetic charge $\eta^2 =
	 \frac{\gamma^2}{4 e^2}$ (cf. eq. (78)). This corresponds to
	 the solutions found in \cite{27} and \cite{28} within the
	 SU(2) theory (due to our different choice of the su(2)-basis
	 one has to replace their $e^2$ by our $4e^2$).

	 \subsubsection*{Ib. The False-Higgs-Vacuum Solutions}

	 Putting $Q^a = 0$ (but keeping $F \neq 0)$, and choosing $w =
	 0$ or $w = 1$ in eq. (55) so that $\vec{B} = 0$ (cf. eqs.
         (57), (58)), we easily find the Reissner-Nordstr\"{o}m solution
	 with a non-vanishing cosmological constant.
	 An example of such a solution is in fact found in Section
	 4.2 $(q_3 = 0, \epsilon = 1)$.

	 \subsubsection*{II Recent Numerical Solutions}

	 Let us set everywhere in a static spherically symmetric
	 spacetime

	 \begin{description}

	 \item (i) $Q_a = 0$, but in general
	 $$
	 V(Q) = \frac{1}{8} k F^4 \neq 0;
	 $$

	 \item (ii) the fields $W^8_0, W^8_A$ and $E_8$ as in eq. (72), (73)
	 and at the internal infinity, $E_8$ and $B_8$ as in eq. (63);

	 \item (iii) $W^a_1 = 0$, $\vec{W}_A$ given by eq. (55) with $w =
	 w(r)$;

	 \item (iv) the metric in the following form (often used for
	 spherically symmetric spacetimes):

	 \bea
	 ds^2 = & - & \left(1 - \frac{2 \gamma^2 m(r)}{r} \right)
	 \sigma^2 (r) dt^2 + \left (1 - \frac{2 \gamma^2 m(r)}{r}
	 \right)^{-1} dr^2 \nonumber \\
	 &+ & r^2 (d \theta^2 + \sin^2 \theta d \varphi^2).
	 \eea

	 \end{description}

	 The Higgs field equation (4) is identically satisfied by this
	 ansatz, the source term in eq. (3) vanishes and the Higgs
	 field contribution to the stress tensor (6) is just the term
	 $ - \frac{1}{4\pi} g^{\mu \nu} V$, which can be interpreted as
	 a cosmological constant. The fields now correspond precisely
	 to the well-known ansatz (see, e.g. \cite{20}) that was
	 recently used also in \cite{23} (with vanishing $B_8$), and
	 we can use the spacetime solutions found in \cite{23} by
	 numerical methods. They are labelled by $r_H$, by a parameter
	 $\alpha$, and by a positive integer $n$. The condition of
	 extremality determines $\alpha$, and it is easy to see that
	 it is equivalent to our eq. (71) in which ${\cal Q} = 0$ and
	 \be
	 \Lambda = - \frac{1}{4} k \gamma^2 F^4.
	 \ee
	 For a more detailed discussion of the properties of these
	 solutions we refer to \cite{23}.

	 Let us just note that one solution at the internal infinity
	 corresponds here to a number of spacetime solutions (with different
	 $n$).

	 {\bf Acknowledgements}. The authors appreciate the
	 discussions with D. Maison, N. Straumann and M. Volkov. J.B.
	 acknowledges the support of the Tomalla Foundation and of the
	 grants GACR-0503 and GAUK-318 of the Czech Republic and
	 Charles University. A.H. acknowledges the support of the
	 Swiss National Science Foundation.

	 \section*{Appendix}

	 Here, we analyze the range of the parameters describing the
	 internal-infinity solutions on the four-dimensional
	 centralizer given in Section 5.1B. That is $Q_8 = 0$, but in
	 general all other fields $\vec{Q}, \vec{B}, E_8,
	 B_8$ are non-vanishing. $\vec{Q}$ and $\vec{B}$ are
	 given by eqs. (56), (57) and (58).

	 First, as a consequence of eqs. (58), (66) and (69) we find
	 the relation
	 $$
	 \left ( \frac{\gamma}{r_H} \right)^2 = \frac{1}{4} \left [ k
	 (\gamma F)^2 + (16 e^2-k) (\gamma {\cal Q})^2 \right]. \eqno(A1)
	 $$
	 Recall (cf. eq. (68) and the discussion following eq. (69))
	 that we assume
	 $$
	 {\cal Q}^2 < F^2. \eqno(A2)
	 $$
	 Further two relations are given by eqs. (70) and (71).

	 Let us discuss some consequences of the relations (A1), (A2),
	 (70) and (71). It is easy to see that the compatibility of
	 eqs. (71) and (A1) requires that
	 $$
	 (16 e^2 - k )(\gamma {\cal Q})^4 - (16 e^2 - k) (\gamma {\cal Q}
	 )^2 + 4 \gamma^2 \Lambda \nonumber \\
	 + k \left[ (\gamma F)^4 - (\gamma F)^2 \right] + \frac{4
	 \gamma^4}{e^2} (E_8^2 + B^2_8) = 0. \eqno(A3)
	 $$

	 Combining eq. (A3) with the condition that the r.h.s. of eq.
	 (70) has to be positive, we obtain the inequality
	 $$
	 \left [2 (\gamma {\cal Q})^2 - 1 \right] \left \{ 16 e^2
	 (\gamma {\cal Q})^2 + k \left[(\gamma F)^2 - (\gamma {\cal
	 Q})^2 \right] \right\}
	 $$
	 $$
	 + \frac{8 \gamma^4}{e^2} (E^2_8 + B^2_8) \geq 0 \eqno(A4)
	 $$
	 which does not involve the cosmological constant. Conversely,
	 if (A3), (A4) and (A2) are satisfied, then (71), (A1) and the
	 positivity of the r.h.s. of (70) are guaranteed. Hence, the
	 possible values of the parameters ${\cal Q}, E_8$ and $B_8$ are
	 determined by relations (A2), (A3) and (A4); the horizon
	 radius is given by eq. (A1).

	 As in \cite{19}, it is useful to discuss separately
	 three cases: (i) $16 e^2 - k >0$, (ii) $16 e^2 - k = 0$, and (iii)
	 $16 e^2 - k < 0$. Since for $E_8 = B_8 = 0$ our equations go
	 over to those of \cite{19} -- except that one has to replace
	 $e^2$ in \cite{19} by $4e^2$ here -- we discover admissible
	 values of ${\cal Q}$ in all three cases. However, more
	 possibilities arise now as in general $E_8 \neq 0$ and $B_8
	 \neq 0$. A detailed discussion will be given elsewhere.

   \end{document}